\documentclass[conference]{IEEEtran}
\IEEEoverridecommandlockouts
\usepackage{cite}
\usepackage{amsmath,amssymb,amsfonts}
\usepackage{algorithmic}
\usepackage{graphicx}
\usepackage{textcomp}
\usepackage{xcolor}
\usepackage{multirow}
\usepackage{makecell}
\usepackage{mathtools}
\DeclarePairedDelimiter\abs{\lvert}{\rvert}

\begin{document}

\title{
Oscillation Analysis and Damping Control for a Proposed North American AC-DC Macrogrid

\thanks{This work is funded by the U.S. Department of Energy (DOE) Office of Electricity Advanced Grid Modelling (AGM) Program. Pacific Northwest National Laboratory (PNNL) is operated by Battelle for the DOE under Contract DOE-AC05-76RL01830.}
}

\author{\IEEEauthorblockN{ \vspace{-0.7cm}\\Kaustav Chatterjee, Sameer Nekkalapu, Antos Varghese, Marcelo Elizondo, Quan Nguyen, Xiaoyuan Fan 
}
\IEEEauthorblockA{\textit{Pacific Northwest National Laboratory} \\ Richland, WA 99354, USA }
}

\maketitle  

\begin{abstract}
In recent years, several studies—conducted by both industry and U.S. Department of Energy (DOE)-funded initiatives—have proposed linking North America's Eastern and Western Interconnections (EI and WI) through a multiterminal DC (MTDC) macrogrid. These studies have explored the advantages and opportunities of the proposed configuration from the perspectives of capacity sharing and frequency support. However, the potential challenges of small-signal stability arising from this interconnection have not been thoroughly examined. To address this gap, detailed model-based simulation studies are performed in this paper to assess the risks of poorly damped inter-area oscillations in the proposed macrogrid. A custom-built dynamic model of the MTDC system is developed and integrated with industry-grade models of the EI and WI, incorporating high levels of inverter-based energy resources. Through model-based oscillation analysis, potential shifts in inter-area modes—for both EI and WI, resulting from the MTDC integration are characterized, and modes with inadequate damping are identified. Furthermore, to mitigate the risks of unstable oscillations, supplementary damping controllers are designed for the MTDC system, leveraging wide-area feedback to modulate active power set points at selected converter stations. A frequency scanning approach is employed for data-driven model linearization and controller synthesis. The damping performance is evaluated under the designed operating conditions and selected contingency scenarios.
\end{abstract}

\section{Introduction}

Ensuring flexibility is becoming increasingly crucial as power systems face greater variability in demand and load patterns \cite{chen_ac_dc, hvdc_pnnl_report}. In this context, High Voltage Direct Current (HVDC) transmission offers a compelling solution, enabling the connection of asynchronous AC systems and facilitating enhanced capacity sharing across interconnections \cite{chen_ac_dc}. 
While most HVDC projects in the United States (US) and globally are currently limited to back-to-back or point-to-point links, there is growing interest in exploring multiterminal DC (MTDC) configurations \cite{doe_mtdc, mtdc_review}. In the past decade, several MTDC projects using Voltage Source Converter (VSC) technology have been deployed in China and parts of Europe \cite{doe_mtdc}. MTDC grids provide significant advantages over traditional two-terminal DC links, including increased redundancy and resilience to single-point failures, improved control over power distribution on AC networks, reduced reserve capacity requirements, and lower lifetime costs \cite{nrc_mtdc, mtdc_review, banerjee_mtdc_dc}. This interest is mirrored in the US, as evidenced by the Midcontinent Independent System Operator’s (MISO’s) proposal for a continent-scale MTDC macrogrid linking the Eastern and Western Interconnections (EI and WI) \cite{miso_macrogrid_report}. 

The 2014 MISO transmission expansion plan featured comprehensive techno-economic studies of the proposed macrogrid, investigating potential DC network topologies and connection points with the EI and WI \cite{miso_macrogrid_report}. Similar analyses for a continent-scale DC macrogrid have also been presented by the Interconnections Seam Study funded by the US Department of Energy (DOE) \cite{Intercon_seams, mcCalley_1}. These studies overwhelmingly indicate high benefit-to-cost ratios in favor of a pan-US macrogrid, demonstrating increased transmission capacity, improved frequency response, and enhanced generation utilization across regions \cite{hvdc_pnnl_report, Intercon_seams, mcCalley_1}. 

That said, with an MTDC macrogrid network interconnecting the EI and WI systems, the operators will face the usual challenges of ensuring stability, particularly those related to poorly-damped inter-area oscillations within each AC interconnection. Power injection from the DC network into the AC (or otherwise) and subsequent rerouting of flows within an AC grid because of DC overlays may modify the system’s eigenvalues and reshape the existing oscillation modes \cite{chatterjee_tpwrd}. As the EI and WI grids evolve with new generation assets and transmission infrastructure such as these MTDC connections, it is essential to reassess system stability. It is worth investigating if there are risks from undamped/poorly-damped oscillations in EI or WI because of these system modifications \cite{mukherjee_iecon, EI_oscillation_renewable, Agrawal2022-vl}. And if so, it also calls for identifying potential mitigation strategies targeting those.

Prior research examining the effects of increasing inverter penetration on system oscillation modes has primarily focused on the EI \cite{you2017impact, EI_oscillation_renewable} and the WI \cite{Elliott2014GM, Agrawal2022-vl, pnnl_pesgm24_task2} as separate entities. These studies have largely concentrated on the proliferation of inverter-based resources (IBRs) on the generation side, and have predominantly considered grid-following inverters while overlooking the potential influence of grid-forming technologies. Notably, there has been a lack of investigation into how inverter-based transmission systems, particularly HVDC, affect inter-area oscillation modes when linking the EI and WI. Although MISO’s macrogrid initiative \cite{miso_macrogrid_report} and several DOE-funded studies \cite{hvdc_pnnl_report, Intercon_seams, mcCalley_1} have comprehensively addressed capacity sharing and macrogrid planning, they have not adequately examined small-signal stability concerns, including the emergence of poorly damped inter-area oscillations. To advance this body of work, the present paper presents a detailed case study of the proposed EI-WI-MTDC interconnection, focusing on the potential adverse impacts on system oscillation dynamics and their mitigation through the strategic design of supplementary damping controllers (SDCs). The SDCs, through fast modulation of active power injections at the MTDC converters, can dampen unstable oscillation modes in interconnected AC systems \cite{mtdc_samp_survey}. A broad range of damping control strategies exists in the literature, from basic linear control designs \cite{nrc_mtdc} to $\mathcal{H}_{\infty}$ and Linear Matrix Inequality (LMI)-based robust controls\cite{banerjee_mtdc_dc}, as well as energy function-based methods and others \cite{mtdc_samp_survey}. A comprehensive survey of these approaches can be found in \cite{mtdc_samp_survey}. 

In this work, a dynamic model of the macrogrid is developed, integrating high-fidelity EI and WI models—sourced from their respective industrial working groups—with a custom MTDC model created at the Pacific Northwest National Laboratory (PNNL). The base models of the EI and WI systems are modified to incorporate high percentages of grid-forming inverter-based generation resources. For oscillation analysis, modal estimation techniques are applied to ringdown data extracted from time-domain disturbance simulations on the system model. The analysis reveals the prospective shifts in the oscillation modes from the base case and flags poorly damaged modes that have the potential to destabilize the system. For damping the critical modes, supplementary damping controllers (SDCs) are designed, which modulate power references at selected MTDC converter stations, utilizing wide-area frequency feedback. The frequency scanning technique \cite{Freq_scan} is employed to estimate a linear model of the system for SDC design. This offline probing method derives a transfer function from measurement data, effectively bypassing the challenges typically associated with model-based linearization in large-scale power systems. The proposed MTDC-SDC framework complements existing HVDC damping controllers in the EI and WI, such as those on the Pacific DC Intertie \cite{pdci}, which were originally designed for legacy inter-area modes without accounting for the macrogrid configuration.


\begin{figure*}[h]
    \centering
    \includegraphics[width=\linewidth]{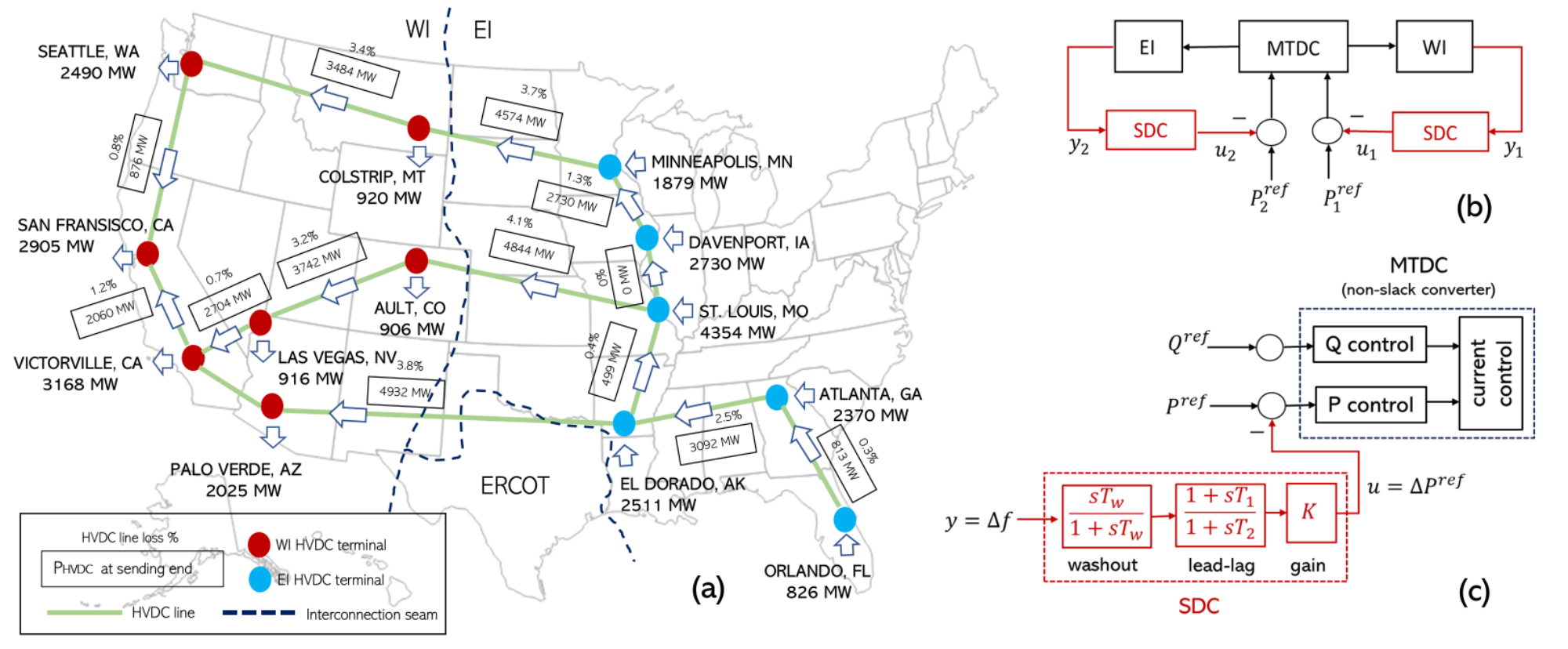}\vspace{-0.3cm}
    \caption{(a) Single-line diagram of MTDC macrogrid showing the scheduled power exchanges between the EI and WI systems, (b) control block diagram for the EI, WI, and MTDC-SDC interaction, and (c) schematic of the SDC (with the constituent blocks) for MTDC power modulation control.}
    \label{fig:macrogrid}
\end{figure*}

\section{System Description and Modeling} \label{sec_sytem}
The system studied is a macrogrid comprising a 13-terminal MTDC network that interconnects full-scale, industry-grade models of the WI and EI, each with high levels of inverter-based generation. The dynamic model of the MTDC is custom-built with SDCs added to selected MTDC converters for power oscillation damping. 

\subsection{MTDC Topology and Parameter Configuration}

The topology of the MTDC network utilized in this work (see, Fig. \ref{fig:macrogrid} (a)) is obtained from the MISO's macrogrid proposal \cite{miso_macrogrid_report}. There are $13$ HVDC substations (or terminals) in the MTDC network -- $6$ in EI and $7$ in WI. The MTDC network is designed to transport approximately $14.4$ GW between EI and WI with an overall power delivery loss of less than $8 \%$ \cite{hvdc_pnnl_report}. There are $14$ DC lines in the topology of which $3$ are connecting the EI and WI, and $11$ overlayed onto the EI and WI systems, $5$ and $6$ respectively (please see, Fig. \ref{fig:macrogrid} (a)). The DC line parameters used in the model -- $2.9$ mH/km, $7.67$ nF/km, and $0.0575$ $\Omega$/mile, are calculated based on a chosen tower and conductor configuration.

\subsection{MTDC Dynamic Model}\label{mtdc_model}

The VSC-based model of the MTDC from \cite{mtdc_model_equations} is considered in this work. One of the $13$ MTDC converters in the system is modeled as the slack which controls the DC-bus voltage and the AC voltage at its location. The remaining $12$ non-slack converters control the active and reactive power injection into the AC grid from their respective terminals. 
The outer voltage and power control loops set the reference points for the inner current control loops in the VSCs. Considering time-scale separation, the inner current control loops are represented by first-order low-pass filters \cite{mtdc_model_equations}. 

\emph{Supplementary Damping Control:} The active power reference $P^{\text{ref}}$ at selected non-slack converters is modulated by a control input $ u = \Delta P^{\text{ref}}$, which is the output of the SDC (see, Fig. \ref{fig:macrogrid} (b)). The schematic in Fig. \ref{fig:macrogrid} (c) details the structure of the SDC and its integration with the existing MTDC controls. The input to the SDC is a wide-area feedback signal, typically a remote bus angle, line flow, or generator speed equivalent, having a high observability of the inter-area modal frequency of interest. The design of the SDC is presented in Section \ref{sec:dc_design}. 

\subsection{WI System Model}

The Western Electricity Coordinating Council (WECC) 2031 planning case \cite{wecc_anchor_data} of WI (23,000+ buses and 2,700+ generators with 163.2 GW generation) is utilized with resource-mix modifications to consider a scenario with $60\%$ IBR penetration. Approximately, 100 GW of conventional generators in the WECC model are replaced with co-located grid-forming inverter (GFM) resources of the same capacities \cite{Xue_automation}. The GFMs, in the study, are represented by the WECC-approved REFGM\_A1 model with the parameter values from \cite{GFMDRP_A}. The protection systems in the GFMs are equipped with active, reactive, and transient fault current limiters.

\subsection{EI System Model}

The Multiregional Modeling Working Group (MMWG) 2020 series 2030 planning case \cite{mmwg_Data} of EI (89,000+ buses and 6,900+ generators with 694.4 GW generation) is used in this study. The generation mix in the EI system is modified to consider a case with $20\%$ IBR penetration. As with the WI case, the fossil fuel-fired generators at selected locations in this case too, are replaced with co-located REGFM\_A1 grid-forming inverters \cite{GFMDRP_A}.

\subsection{Model Integration of WI, EI, and MTDC Systems}

The model of the macrogrid combining the EI, WI, and custom-built MTDC subsystem was integrated for dynamic simulations in the positive-sequence software PSS/E. The custom-built dynamic model of the DC system is written in Fortran and called into PSS/E as a dynamic-linked library (DLL) module \cite{psse_mtdc_implementation}. The in-built solvers in the current versions of PSS/E do not support powerflow solutions for these custom-built DC models. To that end, the sequential AC-DC powerflow algorithm \cite{mtdc_pf, Travis_OSW_Paper} is utilized to obtain a steady-state operating point for dynamic simulations. Events such as generation trips, dynamic brake insertions, etc., are simulated to obtain the time-series data for oscillation analysis and control design. 

\begin{figure}[]
    \centering \vspace{-0.2cm}
    \includegraphics[width= \linewidth]{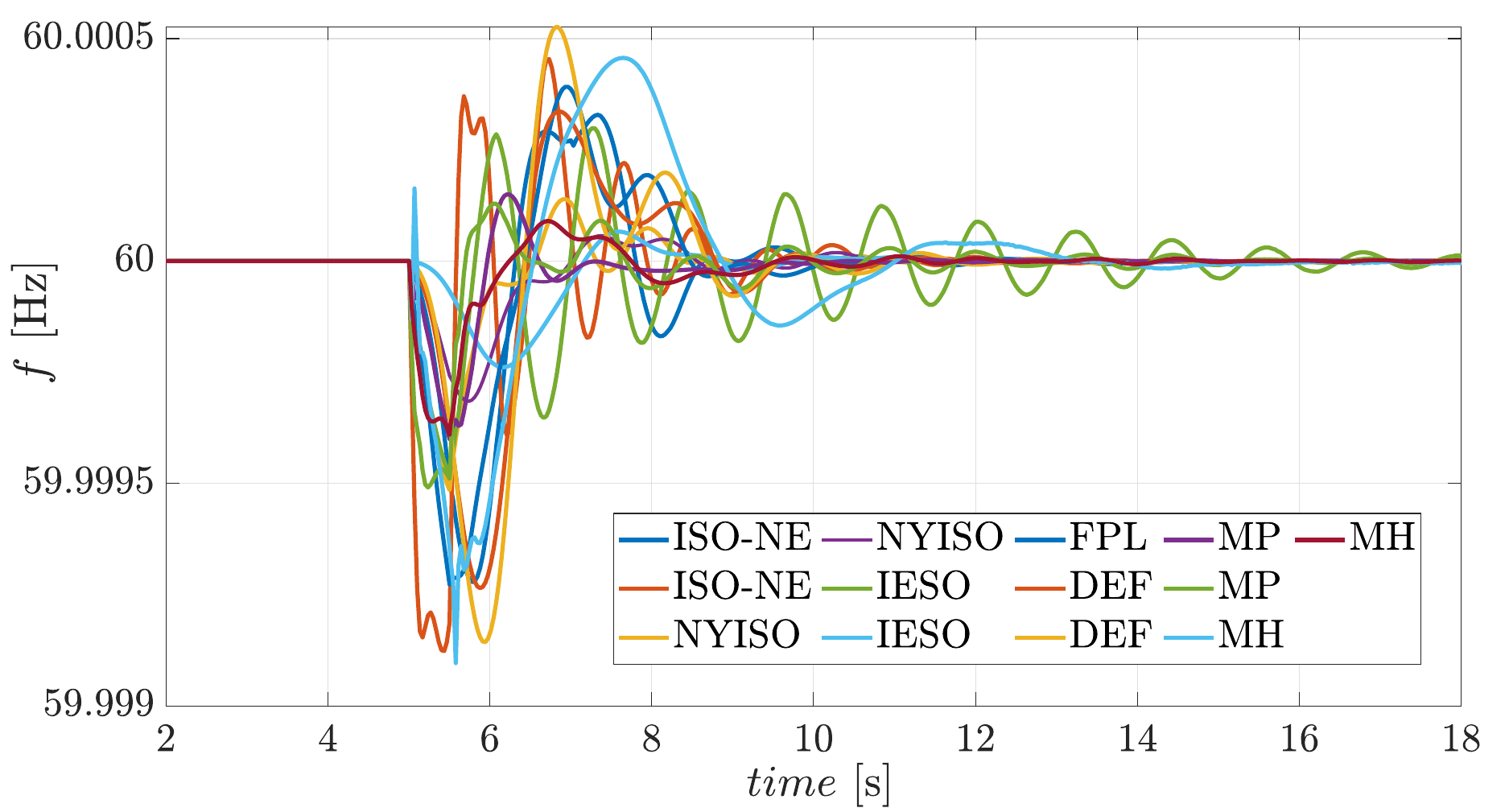}
    \caption{Bus frequencies from different locations in EI following a 1500 MW dynamic brake insertion event in New England.}
    \vspace{-0.4cm}
    \label{cjo_data}
\end{figure}

\begin{figure}[]
    \centering \vspace{-0.2cm}
    \includegraphics[width= \linewidth]{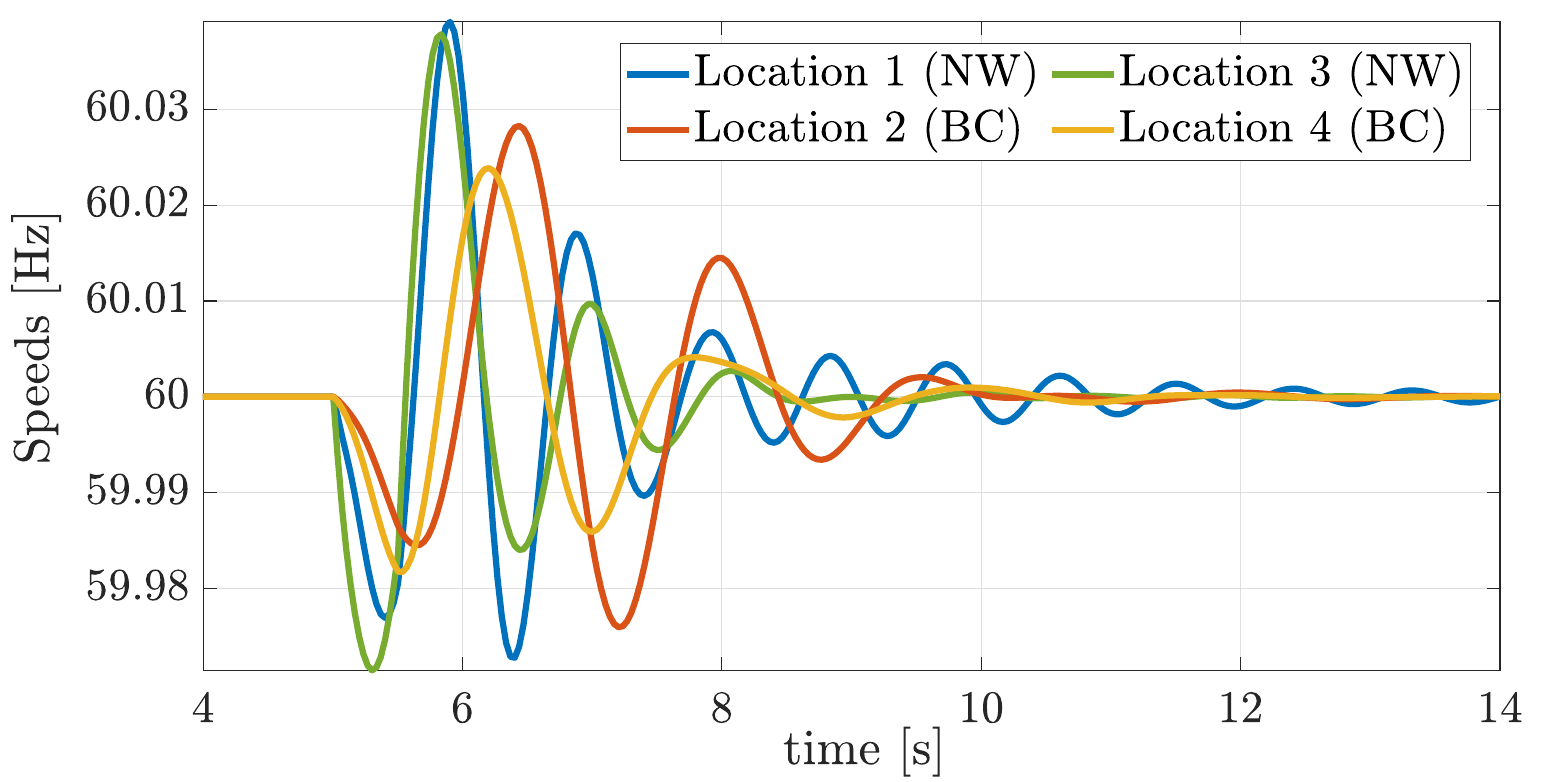}
    \caption{Bus frequencies from different locations in WI following a 1200 MW dynamic brake insertion event in Washington.}
    \vspace{-0.4cm}
    \label{cjo_data}
\end{figure}

\begin{figure*}
    \centering
    \includegraphics[width=\linewidth]{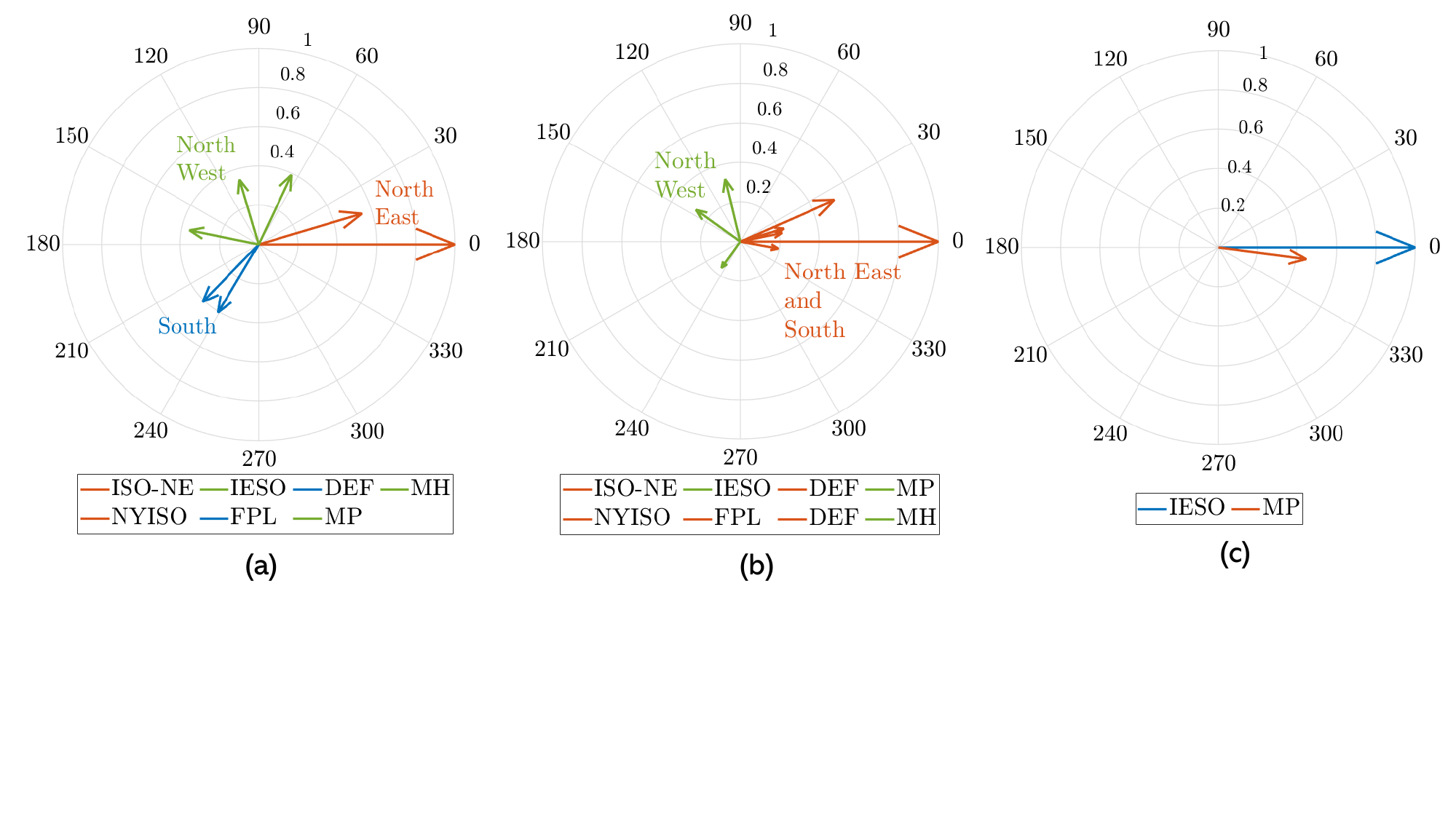}
    \vspace{-3.5cm}
    \caption{Modeshapes of (a) EI Mode 1, (b) EI Mode 2, and (c) EI Mode 3, estimated from the disturbance data. }
    \label{fig:EImodeshape}
    \vspace{-0.2cm}
\end{figure*}

\section{Inter-area Oscillation Modes in the MTDC-Interconnected EI and WI Systems}
This section presents the results of a model-based analysis aimed at identifying newly emergent and shifted inter-area oscillation modes in both the EI and WI systems. These changes arise from modifications to the system configuration, specifically: (1) the replacement of synchronous generators with grid-forming IBRs, and (2) the addition of MTDC transmission interlinking the EI and WI. In performing these analyses, the SDCs at the MTDC converters are disabled to capture the modal characteristics corresponding to the worst-case damping behavior. Once the critical modes and their damping ratios are identified, SDCs will be designed and introduced in the following section to improve the damping of those modes.

As a first step, multiple disturbance events are simulated across various regions within each interconnection, and the resulting time-domain response data is collected from PSSE for analysis. The time series data is pre-processed through appropriate filtering and detrending before modal estimation is performed. To estimate the oscillation modes, two ringdown analysis techniques are employed: (1) the Prony method \cite{multi_prony} and (2) the Matrix Pencil method \cite{matrix_pencil}. Depending on the relative modal observability, suitable signals for mode estimation include real power flows, generator rotor speeds, bus voltage angles, and bus frequencies obtained from the disturbance events. In this study, generator speed measurements resulting from dynamic brake insertion events are used. During such an event, a large resistive load (typically 1000–1500 MW) is temporarily inserted at a selected location for a short duration (approximately 0.3–0.5 seconds) and then removed. This perturbation excites the system’s natural oscillatory modes while allowing it to return to its pre-disturbance steady state. For the EI, the brake events are simulated in the following areas: New York, Minnesota, Virginia, Tennessee, and Florida. Similarly, for the WI, the following locations were chosen for brake event simulations: Washington, California, Arizona, and Montana. 

\begin{figure}[h]
    \centering \vspace{-0.25cm}
    \includegraphics[width=\linewidth]{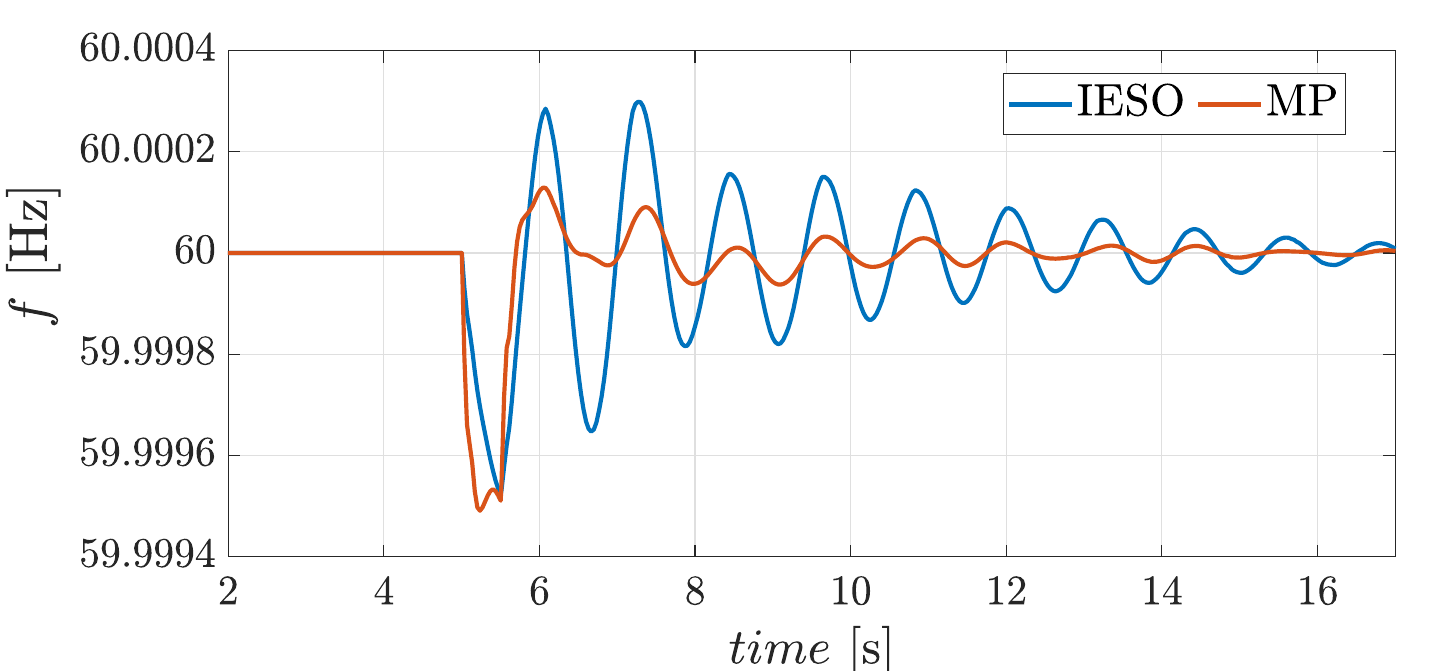}
    \caption{Bus frequencies from IESO and MP, illustrating EI Mode 3 as excited by multiple dynamic brake insertion events in the EI system.}
    \vspace{-0.3cm}
    \label{eimode3_plot}
\end{figure}

\begin{figure*}
    \centering
    \includegraphics[width=\linewidth]{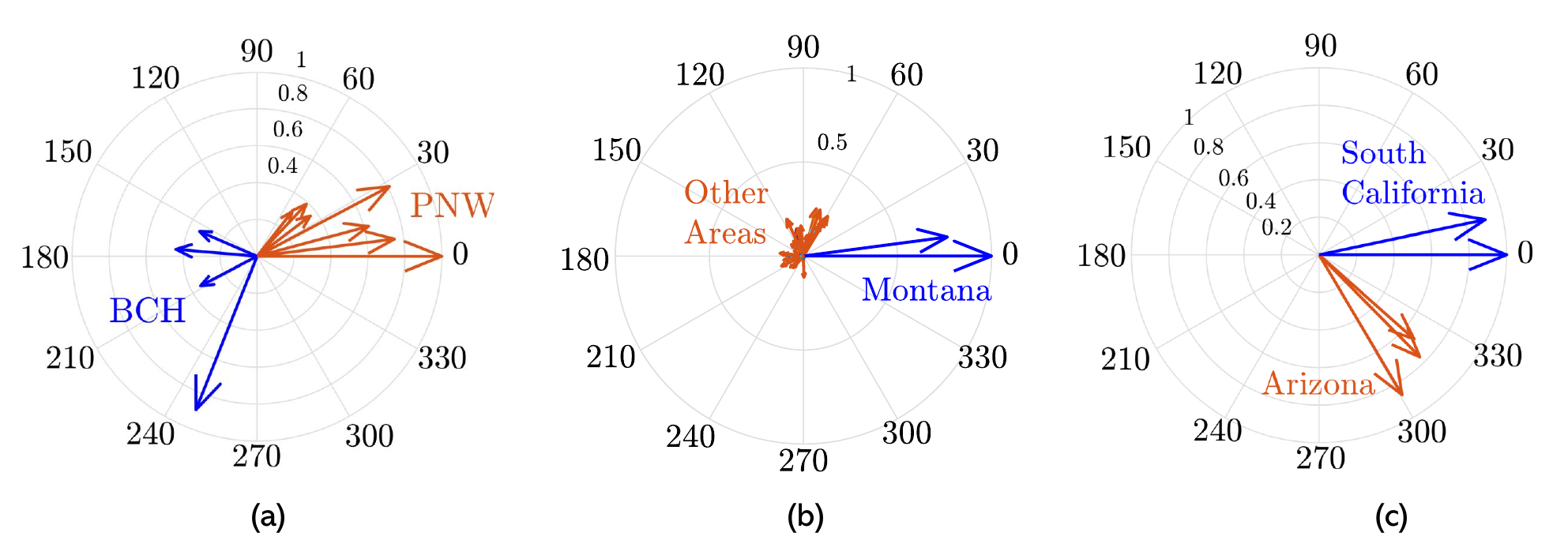}
    \vspace{-0.8cm}
    \caption{Modeshapes of (a) WI Mode 1, (b) WI Mode 2, and (c) WI Mode 3, estimated from the disturbance data. }
    \label{fig:WImodeshape}
    \vspace{-0.2cm}
\end{figure*}

\subsection{Oscillation Analysis for EI}
The analysis of disturbance data from brake insertion events in the EI revealed the presence of two prominent system-wide inter-area modes, referred to as EI Mode 1 and EI Mode 2, as described below: 
\begin{enumerate}
\item \emph{EI Mode 1:} A $0.24$ Hz mode was observed, characterized by three coherent generator groups oscillating against one another: (a) the Northeast (NE) group, comprising generators in New York (NYISO) and New England (ISO-NE); (b) the Northwest (NW) group, including Manitoba Hydro (MH), Minnesota Power (MP), and parts of Ontario (IESO); and (c) the Southern (S) group, represented by generators in Florida such as Duke Energy Florida (DEF) and Florida Power \& Light (FPL). The shape of this mode is illustrated in Fig. \ref{fig:EImodeshape} (a).
\item \emph{EI Mode 2:} A second mode, slightly different in shape from Mode 1, at approximately $0.35$ Hz was observed. The shape of this mode is shown in Fig.  \ref{fig:EImodeshape} (b). In this mode, the generators from Northeastern and Southern EI as a coherent group, oscillate against the generator group in Northwest EI.
\end{enumerate}
Both modes were found to be relatively well-damped under nominal operating conditions—i.e., with no contingencies and all system elements in service—with damping ratios on the order of 12–15\%. These modes exhibit similar characteristics to the NE–NW–S and NW–S inter-area modes identified in the 2019 analysis of the EI base case (prior to any system modifications such as the addition of grid-forming IBRs and MTDC links), as reported in \cite{nerc_oscill_report}. Notably, the 0.16–0.22 Hz North–South (N–S) mode observed in the 2019 base case was not detected in the present analysis. Furthermore, the system modifications introduce a subtle reshaping of the NW–S mode, resulting in participation from the Northeast region in coherence with the Southern generators—manifesting as EI Mode 2 in our study.

However, two new regional inter-area modes emerged from the system modifications that were not observed previously in the study of the base case. The locational observability of these coincided both with the introduction of grid-forming IBRs in those regions as well as with the presence of MTDC terminals nearby. For instance, 
\begin{enumerate}
    \item \emph{EI Mode 3:} A regional mode at approximately 0.84 Hz was observed in the northern part of the system, characterized by generators in IESO (Ontario) and MP (Minnesota) oscillating against the rest of the EI (see Fig. \ref{eimode3_plot}). The shape estimated from data is shown in \ref{fig:EImodeshape} (c). This mode, excited by a brake event near Minneapolis, exhibited low damping with a damping ratio of 3–4$\%$ and a settling time of approximately 15 s.
    \item \emph{EI Mode 4:} A $0.60$ Hz regional mode was observed in Florida, wherein generators and GFMs in FPL and DEF were seen to oscillate against each other. The mode was observed to be reasonably well-damped with a damping ratio of around 12$\%$.
\end{enumerate}
Based on this analysis, EI Mode 3 stands out due to its low damping and proximity to MTDC terminals, indicating heightened vulnerability from system modifications and underscoring the need for targeted control action.

\subsection{Oscillation Analysis for WI}
The analysis of time-series data from disturbance events in the Western Interconnection (WI) reveals the following. Due to the large-scale replacement of synchronous generators—approximately 60\% in WI, compared to only 20\% in the Eastern Interconnection (EI)—the characteristic system-wide inter-area oscillation modes typically observed in the WI base case, such as the North–South and East–West modes \cite{nerc_oscill_report}, were not present in this case. However, three regional inter-area modes were identified, each confined to a specific region within the WI system. With the retirement of fossil fuel-fired generators, these modes were primarily driven by the remaining hydro and geothermal generation. The characteristics of these modes, designated as WI Mode 1, WI Mode 2, and WI Mode 3, are discussed below.
\begin{enumerate}
    \item \emph{WI Mode 1:} An inter-area mode, approximately 0.74 Hz, was observed in the Pacific Northwest (PNW) and British Columbia regions, excited by the brake insertion event near the Chief Joseph substation in Washington. This mode is characterized by a large participation of hydrogenerators, with the units from PNW oscillating against those in BC. The mode has a 10\% damping under nominal condition. However, with critical contingencies involving the tie lines interlinking the regions, the damping reduces significantly, highlighting the need for additional control actions to mitigate the risks. 
    \item \emph{WI Mode 2:} A 0.9 Hz mode was observed in Montana, excited by a disturbance near Colstrip. The mode was well-damped ($\approx$ 18\%) and involved generators in Montana oscillating against the rest of WI. 
    \item \emph{WI Mode 3:} A 0.84 Hz mode was observed in the South, with generators in the Lower Colorado Basin in Arizona and in the Imperial Irrigation District (IID) in Southern California, oscillating against each other and against the rest of the WI system. The mode was well-damped, confined to a limited geographical area, and not observable from other locations. 
\end{enumerate}

The shapes of WI Modes estimated from disturbance events are shown in Fig. \ref{fig:WImodeshape}. Modes 1 and 2 closely resemble the BC and Montana modes identified in the WI base case analysis, respectively \cite{wecc_modereview}. The Mode 3, however, appears to be a newly emergent mode, likely resulting from the splitting of one of the original North–South inter-area modes. A more detailed analysis will be required to fully characterize this mode, which is identified as a topic for future investigation.

In summary, our model-based analysis identifies EI Mode 3 and WI Mode 1 as two critical inter-area modes with low damping, that pose potential stability risks arising from the proposed system modifications. The following sections (Sections IV and V) present appropriate supplementary control strategies leveraging the MTDC system to enhance the damping of these modes.

\section{Supplementary Damping Controller in MTDC} 
\label{sec:dc_design}

In this section, we discuss the design of a supplementary damping controller (SDC)—introduced in Section \ref{mtdc_model}— to modulate the active power reference $P^{\text{ref}}$ at the non-slack converters of the MTDC system using wide-area frequency feedback. The schematic of the SDC is illustrated in Fig. \ref{fig:macrogrid}(c). Signals with high observability of the targeted oscillation mode are ideal candidates for the SDC input. Signal selection is guided by modal analysis performed on time-domain simulation data, calibrated across multiple operating points. The magnitudes of the mode shapes obtained from this analysis serve as quantitative indicators of signal observability. Based on this criterion, wide-area frequency measurements from selected grid locations are chosen as SDC input signals in this study.

\subsection{Structure of the SDC}

From Fig. \ref{fig:macrogrid} (c), observe that the SDC is structured into three functional blocks similar to that of a power system stabilizer (PSS) \cite{kundur}. The \emph{washout block} acts as a high-pass filter to reject the DC offsets and quasi-steady-state slow variations and allows only frequency transients higher than a threshold (ideally, 0.1 Hz in this case -- the lower limit for inter-area modes) to engage the control action. The washout block is followed by a \emph{lead-lag compensator block} and a \emph{gain block}, which add the phase and the gain margins for the desired placement of the closed-loop poles. 
The compensator time constants $T_1$ and $T_2$, and the controller gain $K$, are designed using the pole-placement method. This requires the knowledge of -- (1) the dominant poles or eigenvalues of the open-loop system (i.e., before engaging the SDC), and (2) the desired performance of the closed-loop system (i.e., with SDC) specified in terms of damping ratio $\zeta$  
. 

As is evident, the design approach assumes a linear model of the open-loop system is available. However, for large power systems, like EI or WI, deriving the linearized small-signal model can be extremely challenging. First, the scale of the system (23,000+ buses in WI and 89,000+ buses in EI) impedes the success of accurate model-based linearization and eigenanalysis. Second, the design parameters and internal details of components like converters are the manufacturer's (OEM's) intellectual property which may not be readily available, hindering any realistic chances for model-based linearization. Therefore, any practical linearization effort in a system of this size should be data-driven. To this end, in this work, he frequency scanning technique is used to estimate the transfer function (TF) of the open-loop system numerically from the simulation response. 
The details on the linear model extraction are presented next.

\subsection{Frequency Scanning and Linearized Model Extraction}

The frequency scanning technique \cite{Freq_scan} is widely used to obtain the frequency response of a system by perturbing its input with a small amplitude wide-band probing signal. In this study, for the open-loop system, a multi-sine input signal $u$ with frequencies in the resolution of $0.01$ Hz, in the band $0.05-3$ Hz, of randomly chosen amplitudes and phase offsets is injected to the setpoint $P^{\text{ref}}$ of an MTDC converter with the SDC disengaged. The response observed in the output signal of interest $y$ (i.e., which is a candidate feedback signal for SDC input) is recorded. Between the input-output signal pair $u$ and $y$, a continuous-time transfer function (TF) model of the open-loop system, $G_{\text{OL}}(s) = \frac{y(s)}{u(s)}$, is estimated using the MATLAB Signal Processing Toolbox. In the steps leading to the continuous-time TF estimation, first, the discrete-time state-space model is estimated using the N4SID algorithm \cite{syid}, which is then converted to its continuous-time representation using the $\mathtt{d2c()}$ function, in MATLAB. The poles of $G_{\text{OL}}(s)$ determine the eigenvalues $\lambda_{\text{OL}} = -\sigma_{\text{OL}} \pm j\omega_{\text{OL}}$ of the open-loop system. Errors from overfitting can be avoided by carefully selecting the TF order in the estimation.


\subsection{Design of SDC Parameters} \label{sec:dc_design_leadlag}
Given a performance requirement $\zeta$, the SDC is designed such that the eigenvalues of the closed-loop system, $\lambda_{\text{CL}} = -\sigma_{\text{CL}} \pm j\omega_{\text{CL}}$, achieves $\frac{-\sigma_{\text{CL}}}{\abs{\lambda_{\text{CL}}}} \geqslant \zeta$. In the interest of retaining the modal characteristics, the closed-loop poles are placed such that the oscillation frequency before and after engaging the SDC, ideally, remains unaltered, i.e., $\omega_{\text{CL}}  = \omega_{\text{OL}}$. Following this and the performance requirement on $\zeta$, the $\sigma_{\text{CL}}$ is chosen such that $\sigma_{\text{CL}} > \omega_{\text{OL}}\,\frac{\zeta }{\sqrt{1-\zeta^2}}$.  

Next, let the closed loop TF, after engaging the SDC, between $u$ and $y$ be represented as $G_{\text{CL}}(s) = \frac{G_{\text{OL}}(s)}{1 \,+\, G_{\text{OL}}(s)\, H(s)}$,
where $H(s) = K\, \Big(\frac{1 \, + \, sT_1}{1 \, + \, sT_2}\Big)^m \frac{sT_w}{1\,+\,sT_w}$ is the TF of the SDC and $m$ is the number of cascaded lead-lag compensators in the design. 
With this premise, the SDC is designed as follows.
\begin{enumerate}
    \item First, the desired phase compensation $\angle{ H(s)}$ at the closed-loop pole $\lambda_{\text{CL}}$ is calculated. Note that, since $G_{\text{CL}}(s)$ has a pole at $s = \lambda_{\text{CL}}$, $ H(\lambda_{\text{CL}}) \approx -\frac{1}{G_{\text{OL}}(\lambda_{\text{CL}})}$. Following this, it is checked if $\angle{ H(\lambda_{\text{CL}})}$ is less than the compensation value that can be provided by a single lead-lag block, which for all practical designs is restricted to $50-55^\circ$. 
    \item If $\angle{ H(\lambda_{\text{CL}})}$ is greater than this maximum compensation value, the number of cascaded lead-lag blocks is increased, and $m$ is chosen accordingly.

    \item Next, define $\phi = \frac{\angle{ H(\lambda_{\text{CL}})}}{m}$. Using which, calculate $\alpha = \frac{1 \,- \,\sin\phi}{1 \,+\, \sin{\phi}}$. The lead-lag compensator time-constants, $T_1$ and $T_2$ are designed as, $T_1 \,=\, \frac{1}{\omega_{\text{CL}}\,\sqrt{\alpha}}, \,\,\, \text{and}\, \, \, T_2 \, = \, \alpha\,T_1.$
    
    \item To reject slow variations below $0.1$ Hz, the washout block is designed with time-constant $T_w = 10$ s.

    \item The gain parameter $K$ is designed as the ratio of the desired gain $\abs{H(\lambda_{\text{CL}})}$ and the gain contributed by the washout and the lead-lag blocks at the closed-loop pole. It may be fine-tuned through multiple time-domain simulations to match the settling time requirements. 
\end{enumerate}
The implementation of the MTDC-SDC in the EI-WI macrogrid is discussed in the following section.

\section{Implementation of SDC and Results }

This section presents the implementation details and performance evaluation of the SDCs developed for the MTDC macrogrid. Two actuation points are selected—one in the WI at Seattle and another in the EI at Minneapolis. At each of these locations, the SDC is implemented as a single-input, single-output (SISO) system and is designed using the frequency scanning technique described in Section \ref{sec:dc_design}.

As identified in the oscillation analysis for WI (Section III), a critical 0.74 Hz mode (WI Mode 1) was observed in the British Columbia (BC)–Pacific Northwest (PNW) region, exhibiting strong controllability from Seattle. Accordingly, the MTDC converter at Seattle is selected as the actuation point. A remote bus frequency signal from a generator in the BC region, $f_{\text{BC}}$, having the highest observability of the mode, is used as the feedback signal for the SDC at the Seattle converter. 

Similarly, in the EI system, the poorly damped 0.84 Hz mode (EI Mode 3) is targeted using the MTDC converter at Minneapolis. A remote bus frequency signal from Ontario's IESO region, $f_{\text{ON}}$, with high observability of this mode serves as the feedback input for this SDC.

\begin{figure}
    \centering \vspace{-0.2cm}
    \includegraphics[width=0.49\linewidth]{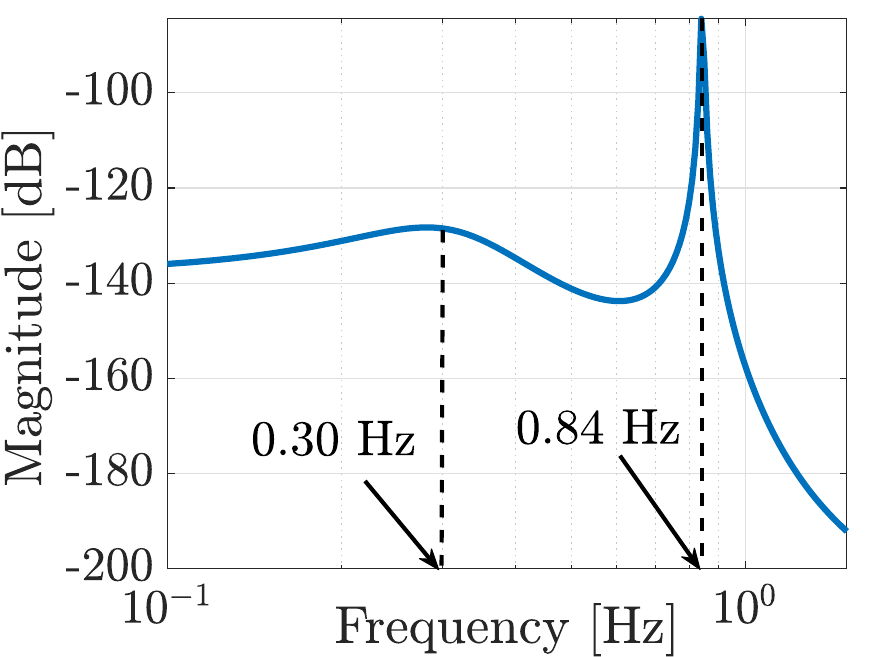}
    \includegraphics[width=0.49\linewidth]{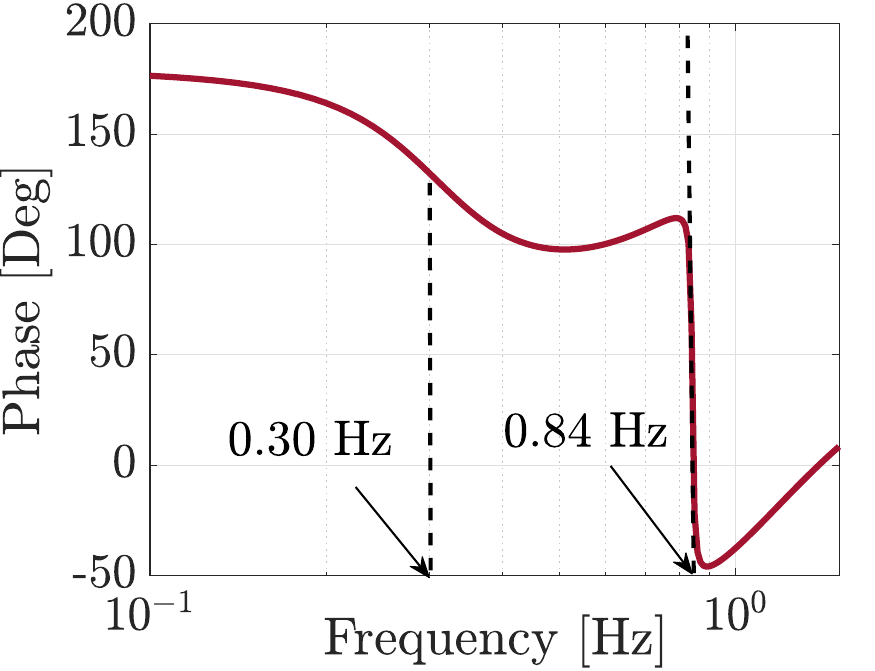}\vspace{-0.2cm}
    \caption{Bode plot of $G_{\text{OL}}^{\text{EI}}(s)$ showing magnitude and phase responses for frequencies in the $0.1-2$ Hz range}
    \label{fig:bode_EI}
    \vspace{-0.0cm}
\end{figure}

\begin{figure}
    \centering \vspace{-0.0cm}
    \includegraphics[width=\linewidth]{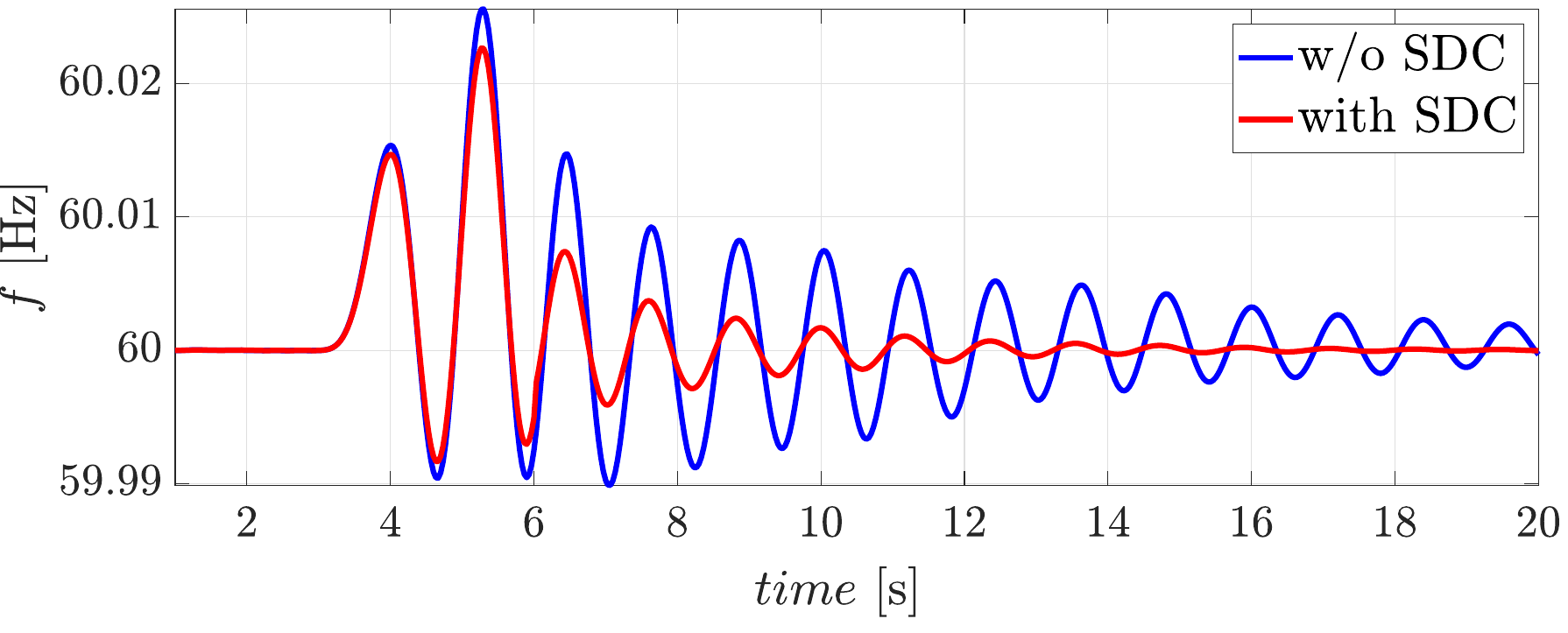}\vspace{-0.2cm}
    \caption{Bus frequency at a location in the EI region with and without the action of the MTDC SDC at Minneapolis.}
    \label{fig:NY_speed}
    \vspace{-0.3cm}
\end{figure}

\subsection{Design and Performance of SDC at Minneapolis in EI}

Using the frequency scanning technique, the open-loop TF, $G_{\text{OL}}^{\text{EI}}(s) = \frac{y_1(s)}{u_1(s)}$, is estimated between the wide-area frequency feedback signal $y_1(s) = \Delta f_{\text{ON}}$ from IESO region and the multi-sine excitation input $u_1(s) = \Delta P^{\text{ref}}_{\text{Minn.}}$ at Minneapolis. The magnitude and phase plots of $G_{\text{OL}}^{\text{EI}}(s)$ are shown in Fig. \ref{fig:bode_EI}. The open-loop TF has two dominant poles -- the one at $0.30$ Hz is well-damped while the other at $0.84$ Hz is poorly-damped. The SDC at Minneapolis is designed to target the $0.84$ Hz mode. The design considers enhancing the damping from $\zeta = 3.4\%$ to $12\%$. Following the steps outlined in Section \ref{sec:dc_design_leadlag}, SDC parameters obtained from the design are: $T_1 = 0.1483$ s, $T_2 = 0.2409$ s, $T_w = 10$ s, and $K = 204$. To verify the performance of the damping control, the system is perturbed by applying a pulse excitation to the MTDC input. The responses in Fig. \ref{fig:NY_speed} illustrate the outcome of the control action. Observe that the SDC while improving the damping for the targeted $0.84$ Hz mode while minimally affecting the already well-damped $0.30$ Hz mode.  

\begin{figure}[t]
    \centering \vspace{-0.2cm}
    \includegraphics[width=0.49\linewidth]{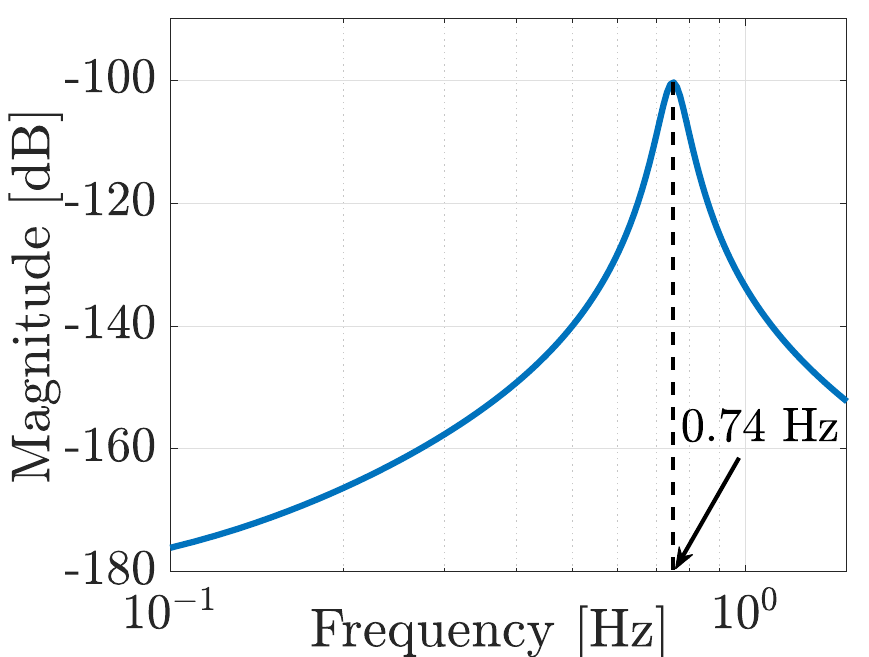}
    \includegraphics[width=0.49\linewidth]{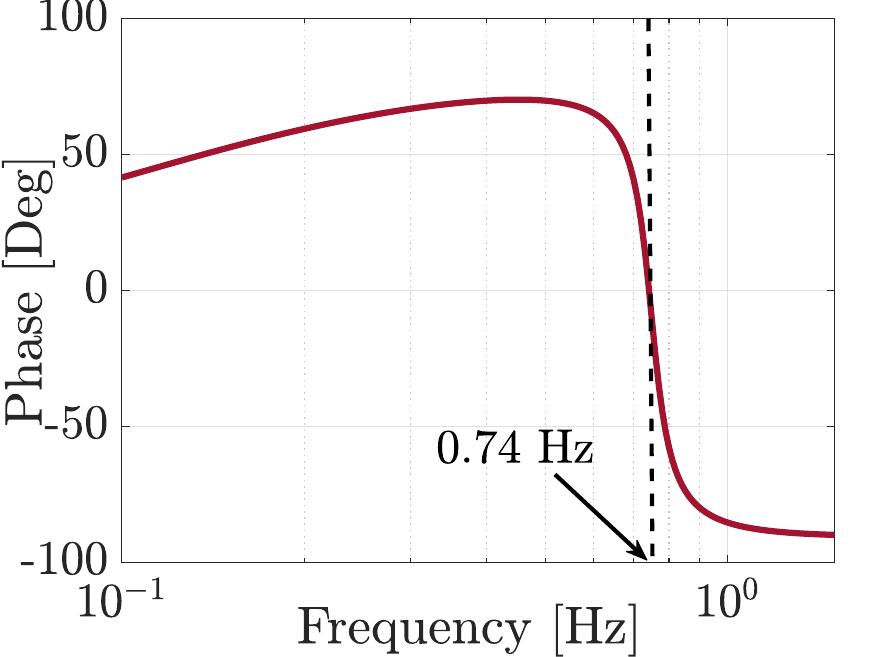}\vspace{-0.1cm}
    \caption{Bode plot of $G_{\text{OL}}^{\text{WI}}(s)$ showing magnitude and phase responses for frequencies in the $0.1-2$ Hz range}
    \label{fig:bode_WI}
\end{figure}

\begin{figure}
    \centering 
    \includegraphics[width=\linewidth]{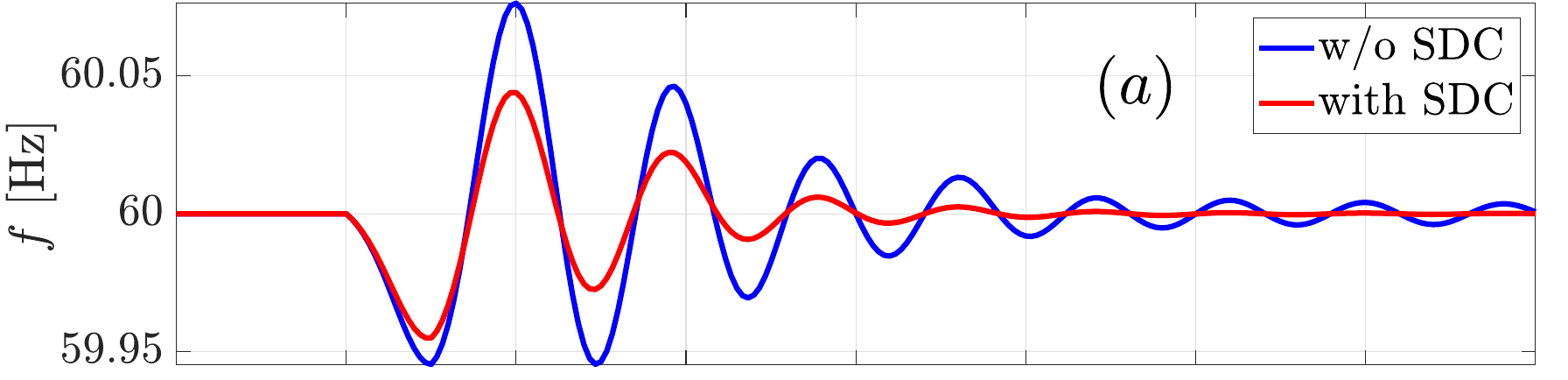}\vspace{0.1cm}
    \includegraphics[width=\linewidth]{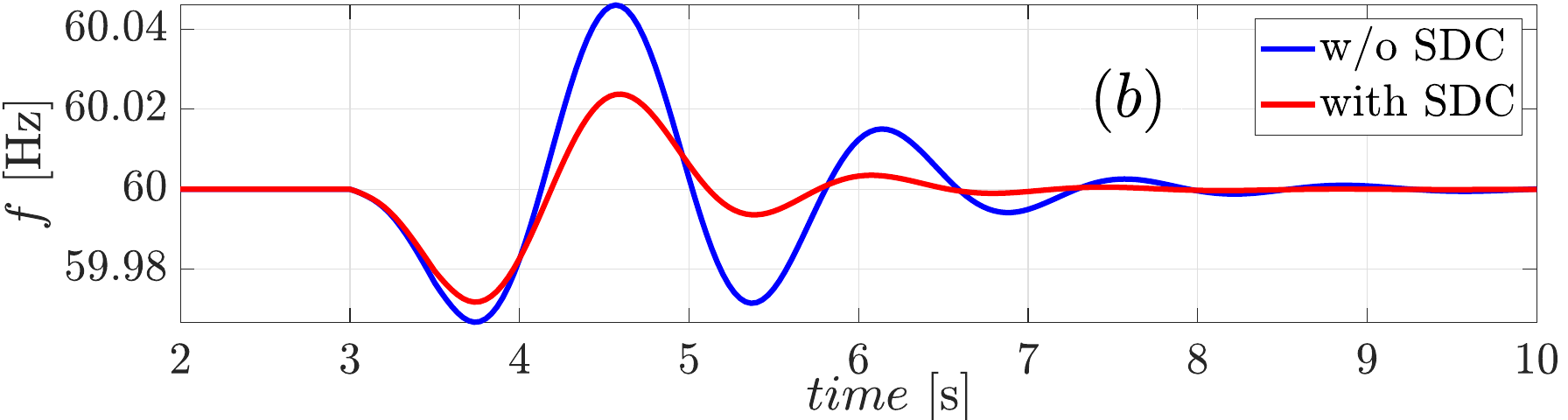}\vspace{-0.2cm}
    \caption{Bus frequencies at two locations in the WI region with and without the action of the MTDC SDC at Seattle.} 
    \label{fig:JOR_speed}
   \vspace{-0.3cm}
\end{figure}

\subsection{Design and Performance of SDC at Seattle in WI}

Using the frequency scanning approach, the open-loop TF,
$G_{\text{OL}}^{\text{WI}}(s) = \frac{y_2(s)}{u_2(s)}$, between the signal pair $u_2(s) = \Delta P^{ref}_{\text{Seattle}}$ and $y_2(s) = \Delta f_{\text{BC}}$ is estimated. The magnitude and phase responses for the TF are shown in Fig. \ref{fig:bode_WI}. As observed from these plots, the open-loop system has a pole at $0.74$ Hz. From the characteristics equation of the estimated TF, the open-loop pole is calculated to be $-0.266 \pm j 4.69$ with $5.6\%$ damping. To achieve a $\zeta > 0.15$ (i.e., $15\%$ damping) without altering its modal frequency, the closed-loop pole is desired to be at $- 0.8 \pm j 4.69$. 
For this design choice, the phase compensation required from the SDC, $\angle{H(s)}$ at $s= \lambda_{\text{CL}}$, is greater than what a single lead-lag block can provide for practical designs. Therefore, instead of one, three lead-lag blocks are cascaded in series (i.e., $m=3$). Next, following the steps in Section \ref{sec:dc_design_leadlag}, the SDC parameters are obtained as  $T_1 = 0.0426$ s, $T_2 = 2.611$ s, $T_w = 10$ s, and $K = 900$. 

For damping performance evaluation, a 1200 MW dynamic bake insertion event is simulated at the Chief Joseph substation of the WI, for 0.5 s. The bus frequencies at a location in the NW region, with and without the SDC, are compared in Fig. \ref{fig:JOR_speed}. It is evident from the plot that engaging SDC enhances the damping of the targeted mode. 


\subsection{Performance of the SDC Under Contingencies} 
\begin{figure}[h]
    \centering 
    \includegraphics[width=\linewidth]{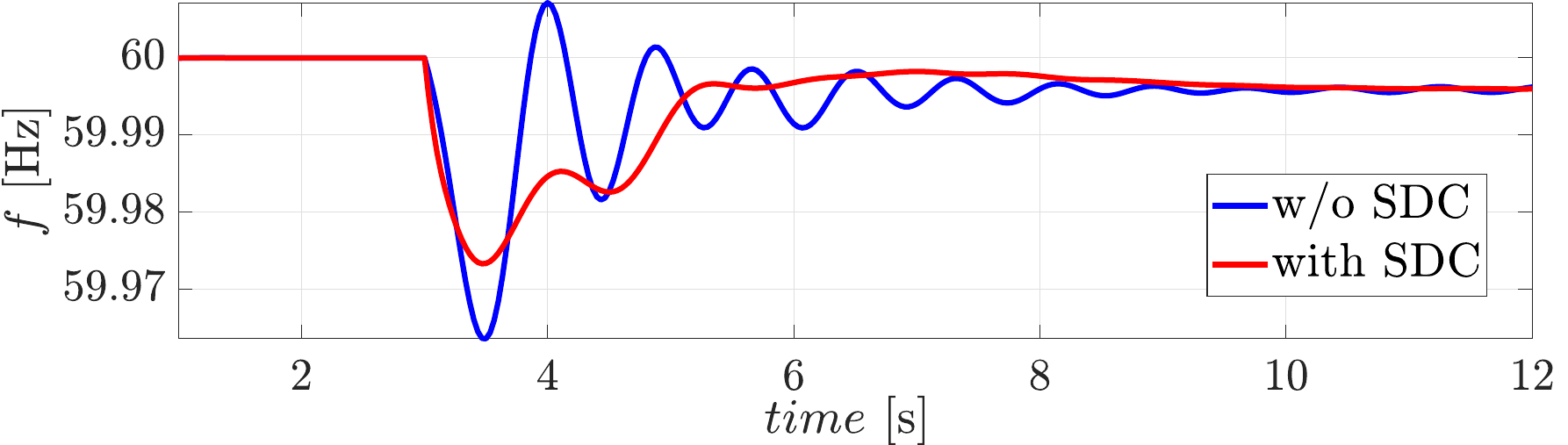}\vspace{-0.2cm}
   \caption{Bus frequency at a location in WI, with and without the action of SDC, for a generation trip resulting in a different post-event operating point.}
    \label{fig:CGS} \vspace{-0.3cm}
\end{figure}

Although SDCs are designed for specific operating conditions, their damping performance is expected to be preserved for minor shifts in the operating points, like those during system contingencies, which can alter system topology, operating points, and modal characteristics. To evaluate this robustness, the performance of the SDC implemented at the Seattle MTDC converter is tested under a representative contingency scenario involving a 1000 MW generator trip in the PNW region of the WI. The bus frequency from a nearby location is monitored and compared for two cases: with the SDC deactivated and with the SDC engaged. The results, shown in Fig. \ref{fig:CGS}, reveal that without the SDC, the 0.74 Hz mode exhibits significantly slower damping and sustained oscillations following the disturbance. In contrast, when the SDC is active, the same mode is rapidly attenuated.


\section{Conclusions and Futurework}
This paper presented a case study of power oscillation damping in a proposed multiterminal DC macrogrid of North American Eastern and Western Interconnections (EI and WI). 
The topology for this MTDC network was obtained from a transmission planning study presented by the MISO, while its dynamic model was custom-built. 
The EI and WI systems were interfaced with high percentages ($20\%$ and $60\%$ respectively) of grid-forming inverter-based generation resources. 
Modal analysis was conducted on ringdown responses extracted from time-domain simulation data. The estimated modal frequencies and mode shapes revealed significant shifts in oscillatory characteristics, as well as the emergence of new modes resulting from the system modifications applied to the baseline EI and WI models. Additionally, the damping ratios of the identified modes were evaluated to determine which modes posed potential stability concerns.
For enhancing the damping of critical inter-area oscillations in each interconnection, a supplementary feedback control strategy modulating the active power points at selected MTDC converter stations was designed. In this study, two converter stations -- Seattle in WI and Minneapolis in EI, were chosen for damping control due to the high controllability of the targeted oscillation modes from these MTDC converter sites. The damping controllers used lead-lag compensator blocks, the parameters of which were designed using linear control. The frequency scanning technique was used to estimate the linearized model of the open-loop system used in control synthesis. The performance of the damping control was tested for the base case system, as well as for the system under stressed conditions with contingencies. Future work will explore multi-objective optimizations for damping controllers with multi-input feedback designs.

\bibliographystyle{ieeetr}
\bibliography{agm}
\end{document}